%% file: poet.tex
\documentclass[runningheads]{llncs}
\usepackage{graphicx}
\usepackage{multirow}

\begin{document}
\title{
POET: A Self-learning Framework for PROFINET Industrial Operations Behaviour
\thanks{Supported by topic Engineering Secure Systems of the Helmholtz Association.}
}
\titlerunning{PROFINET Operations Enumeration and Tracking}
%
\author{
 Ankush Meshram\inst{1}\orcidID{0000-0001-6903-9446} \and
 Markus Karch\inst{2}\orcidID{0000-0002-5683-4499} \and
 Christian Haas\inst{2} \and 
 Jürgen Beyerer\inst{1,2}\orcidID{0000-0003-3556-7181}
}
\authorrunning{
Meshram et al.
}
%
\institute{
 KASTEL Security Research Labs, \\
 Vision and Fusion Laboratory (IES), 
 Karlsruhe Institute of Technology, \\
 76131 Karlsruhe, Germany \\
 \email{ankush.meshram@kit.edu} 
 \and
 Information Management and Production Control, \\
 Fraunhofer Institute of Optronics, System Technologies and Image Exploitation (IOSB),
 76131 Karlsruhe, Germany \\
 \email{\{markus.karch,christian.haas,juergen.beyerer\}@iosb.fraunhofer.de}
}
\maketitle              
\input{content/00-abstract} 
\input{content/01-introduction}

\input{content/02-preliminaries}

\input{content/03-networkoperations} 
\input{content/04-poet}

\input{content/05-anomalydetection} 
\input{content/06-conclusion}
%
%
%

%
%
%
\bibliographystyle{splncs04}
\bibliography{poetbibliography}

\end{document}

%% file: content/00-abstract.tex
\begin{abstract}
Since 2010, multiple cyber incidents on industrial infrastructure, such as \textit{Stuxnet} and \textit{CrashOverride}, have exposed the vulnerability of Industrial Control Systems (ICS) to cyber threats.
The industrial systems are commissioned for longer duration amounting to decades, often resulting in non-compliance to technological advancements in industrial cybersecurity mechanisms. 
The unavailability of network infrastructure information makes designing the security policies or configuring the cybersecurity countermeasures such as Network Intrusion Detection Systems (NIDS) challenging. 
An empirical solution is to self-learn the network infrastructure information of an industrial system from its monitored network traffic to make the network transparent for downstream analyses tasks such as anomaly detection. 
In this work, a \textit{Python}-based industrial communication paradigm-aware framework, named \textsl{PROFINET} Operations Enumeration and Tracking (POET), that enumerates different industrial operations executed in a deterministic order of a \textsl{PROFINET}-based industrial system is reported.
The operation-driving industrial network protocol frames are dissected for enumeration of the operations.
For the requirements of capturing the transitions between industrial operations triggered by the communication events, the Finite State Machines (FSM) are modelled to enumerate the \textsl{PROFINET} operations of the device, connection and system. 
POET extracts the network information from network traffic to instantiate appropriate FSM models (Device, Connection or System) and track the industrial operations. 
It successfully detects and reports the anomalies triggered by a network attack in a miniaturized \textsl{PROFINET}-based industrial system, executed through valid network protocol exchanges and resulting in invalid \textsl{PROFINET} operation transition for the device.

\keywords{Network Security  \and Cyber-Physical System \and Intrusion Detection.}
\end{abstract}

%% file: content/01-introduction.tex
\section{Introduction}

The incremental advancement in technology since steam-powered manufacturing mechanization (the \textit{first} industrial revolution) to the \textit{third} industrial revolution of computer-driven process automation resulted in the \textit{fourth} industrial revolution of cyber-physical systems.
Industrial production systems of the 21st Century are designed for production cost reduction through efficient control of cyber-physical industrial components, realized through the adaptation of Ethernet technology in industrial networking. 
It blurred the separation between the \textsl{office networks} and the \textsl{industrial networks} to allow the personnel in the corporate office access a sensor in the production floor.
The blurring of network separation led to cybersecurity vulnerabilities to industrial production, as demonstrated by recent cyber incidents from \textit{Stuxnet} to \textit{Triton} in the last decades \cite{icsvulnerabilities}. 
The Advanced Persistent Threats (APT) modified the instruction exchanges within industrial processes, utilizing industrial protocols, to cause structural damage to components (\textit{Stuxnet}) or Human, Societal and Environmental (HSE) hazards (\textit{Industroyer/CrashOverride, Triton}).
Continuous monitoring and analysis of industrial communication characteristics is required to detect the initiation of such attacks and reduce the dwell time to accelerate mitigation.

An anomaly-based Network Intrusion Detection System (NIDS) is one of the cybersecurity countermeasures that monitors the network traffic of an industrial production and learns the characteristics from its network infrastructure information to detect the deviations as \textit{anomalies}.
Germany's Federal Office for Information Security (BSI) in its cybersecurity recommendations on production networks \cite{bsireco} outlined anomalies in industrial networks and related categories of feature requirements for anomaly detection systems.
The general requirements category collectively needs the anomaly detection to provide overview of all devices communicating in the network and identify the communication links along with the protocols used.
Another requirement category emphasizes on the ability to detect unusual or exceptional activities in an industrial network, such as identification of new devices in the network, new protocols or changes in protocol among individual components, etc. 

\subsection{Problem Statement}

In the absence of network infrastructure information, such as asset inventory and network policies, of an existing industrial system, designing the security policies or configuring the cybersecurity countermeasures such as NIDS is challenging. 
Insights into the industrial system's operation are required for efficient monitoring and timely incident, \textit{cyber} and \textit{physical}, response.
Interpretation of industrial system operations from its communication network characteristics contributes to being vigilant of cyber threats aimed at industrial process disruption. 
An empirical solution is to self-learn the network infrastructure information (topology, assets and communication links) and the characteristic behaviour from passive monitoring of industrial network traffic, in conjunction with an anomaly detection system to detect anomalies. 
The industrial network's topology, communication relations, assets and protocol data being exchanged during the industrial process operations are the network information to be extracted from the network traffic passively. 
The detection of the different industrial operations from start-up to process data exchange from the monitored network traffic in a systematic way for a self-learning approach is a challenge for \textit{network transparency} (\textbf{Problem 1}). 

In addition, the industrial operations executed in an order create the foundation for process data exchanges realizing the underlying intended process. 
Enumerating these operations through the analysis of multiple protocol communications observed in the traffic and tracking their executions helps to define the industrial system's operation behaviour. 
Monitoring the valid operations of devices, communication links or the industrial system would detect the adversarial actions in the context of protocol specifications of employed industrial networking technology such as \textsl{PROcess FIeld NETwork (PROFINET)} \cite{pi2018}. 
The representation and enumeration of a \textsl{PROFINET} system's operations from monitoring the multiple protocol communications in the network traffic for self-learning its \textit{industrial operation behaviour} is another challenge (\textbf{Problem 2}). 

There are multiple research works in the literature and commercial NIDS solutions that model the message exchanges of industrial network protocols based on corresponding protocol specifications, and the deviations are classified as anomalies \cite{KippeKarch2021}.  
In particular, Snort rules for \textsl{MODBUS} \cite{snortmodbus}, Bro rules for \textsl{DNP3} \cite{brodnp3} and specification-based IDS for \textsl{GOOSE} \cite{goose} check for the validity of packet fields and communication exchanges. 
However, the effect of protocol exchange on the industrial system is not modeled. 
A valid protocol exchange could have adverse effect on the industrial operations which hasn't been modeled in any of the reported works.

\subsection{Proposed Solution}

Different industrial networking operations of a \textsl{PROFINET}-based industrial system executed with different industrial protocols are mapped to corresponding industrial operations from the start up to the process data exchange operation. 
\textsl{PROFINET}'s specifications are followed to correctly map network protocols to detect networking operations and their constituent stages in a Python-based framework that passively captures the network traffic, and extracts relevant information to make the industrial network transparent for analysis.
The network information made available through the developed network transparency solution are utilized to enumerate different industrial operations whenever they occur. 
An industrial system's operation behaviour is considered at device-level, connection-level and system-level to track operational state changes in devices, established process exchange communication links and the overall system. 
Graph-represented Finite State Machines (FSM) are conceptualized for each device, connection and industrial system, where nodes represent the stages of industrial operations and the edges are the transitions that are triggered by the events observed in the extracted network information from the traffic. 
A \textit{Python}-based framework, named \textsl{PROFINET} Operations Enumeration and Tracking (POET), that extracts the network information from \textsl{PROFINET}-based industrial system's network traffic to instantiate appropriate FSM models (Device, Connection or System) and track the industrial operations is developed. 
On a miniaturized \textsl{PROFINET}-based industrial demonstrator (\textit{Festo Demonstrator}), the POET is successfully employed to detect anomalies triggered by a network attack targeted at an industrial component, executed through valid \textsl{PROFINET Discovery and Configuration Protocol (PN-DCP)} exchanges and resulting in invalid \textsl{PROFINET} operation transition for the device.

In the next section, a brief overview of \textsl{PROFINET} technology followed with information on miniaturized \textsl{PROFINET}-based industrial system under consideration and simulated industrial attack scenario is provided. 
In Section~\ref{sec:profinetoperation}, the enumeration of different networking operations executed through different network protocols in \textsl{PROFINET} systems is provided. 
The proposed framework to enumerate and track the \textsl{PROFINET} operations from the analysis of traffic data is summarized in Section~\ref{sec:enumerateandtrack}.
The Section~\ref{sec:poetusage} presents the framework's usage for anomaly detection along with brief discussion, and conclusion in Section~\ref{sec:conclusion}.

%% file: content/02-preliminaries.tex
\section{Preliminaries}

\subsection{PROFINET}

Proprietary fieldbus protocols were developed to satisfy the strict requirements for real-time data transmission and deterministic communication for industrial network operations such as \textsl{PROFIBUS}, \textsl{Modbus}, etc.
\textsl{PROFINET} is the result of adapting \textsl{PROFIBUS} to real-time technology and standardized in IEC 61158 \& IEC 61784. 
\textsl{PROFINET} has $18 \%$ market share in the industrial networks that are installed globally in 2021 as compared to $17 \%$ \textsl{EtherNet/IP} \cite{hms}. 
Additionally, \textsl{PROFINET} is the leader of Industrial Ethernet technology in the European market which concluded its selection as the industrial system under consideration for the presented work. 

There are two real-time properties of \textsl{PROFINET} communication: (a) non-synchronized real-time communication (RT), and (b) synchronized real-time communication (IRT).
Within \textsl{PROFINET}, process data and alarms are transmitted with RT communication with bus cycle times in the range of $50-100\,ms$.
Isochronous data transfer with IRT communication is used in applications such as motion control requiring bus cycle times in range of microseconds, $<1\,ms$.
In addition, \textsl{PROFINET} defines different classes of components characterized by their functionality and participation at different stages of industrial communication - IO Controller, IO Supervisor and IO Device. 
An IO Controller is the component with \textit{master} functionality that executes the automation program, typically a Programmable Logic Controller (PLC).
It participates in parametrization, cyclic/acyclic data exchange and alarm processing with connected field devices.
An IO Supervisor is used for the commissioning and diagnostic purposes, generally a programming device, personal computer or Human Machine Interface (HMI). 
An IO Device is a field device in the vicinity of process with \textit{slave} functionality that sends process data and critical statuses (alarms \& diagnostics) to connected IO Controller(s) via \textsl{PROFINET}.
The transmission of data from an IO Controller/Supervisor to an IO Device is designated as \emph{output data} whereas IO Device to IO Controller/Supervisor is \emph{input data}.
\textsl{PROFINET} utilizes the provider-consumer model of communication for I/O data exchange between controllers and devices, as well as parametrization and diagnosis information exchange between supervisors and devices.

An IO Device comprises of an Ethernet interface for communication and physical/virtual modules to handle the process data traffic. 
The device model of an IO Device consists of \textit{slots}, \textit{subslots}, \textit{modules}, \textit{submodules} and \textit{channels}. 
The slot and subslot designates the insert slot of a module and submodules in an IO field device, respectively. 
The module provides the structuring, and contains at least one submodule which always holds the process data with status information. 
The data within the submodule is addressed using an index. 
Cyclic IO data in submodule are accessed through slot/subslot combinations, whereas, acyclic read/write services utilize slot, subslot and index. 


\subsection{System Under Consideration}
\label{sec:suc}

The quality and characteristics of dataset employed in the development of ICS cyber threat detection methods play an important role in driving the Industrial Cybersecurity research. 
In the quest for finding the solutions to the aforementioned challenges, a \textsl{PROFINET}-based industrial system is used for developing and evaluating the proposed solutions.
A \textsl{PROFINET}-based scaled-down industrial system with real industrial components and fully functional networking infrastructure, labeled as \textsl{Festo Demonstrator} 
is employed for the reported analysis. 
The network attacks targeted at the \textsl{Festo Demonstrator}'s underlying network and process operations are scripted, executed and resulting anomalies are passively captured from the network traffic. 
A systematic Python-based framework passively captures the network traffic from \textsl{PROFINET}-based system and extracts relevant information to make the industrial network transparent for downstream analyses such as anomaly detection. \\
\begin{figure}
	\centering
	\includegraphics[width=0.85\linewidth]{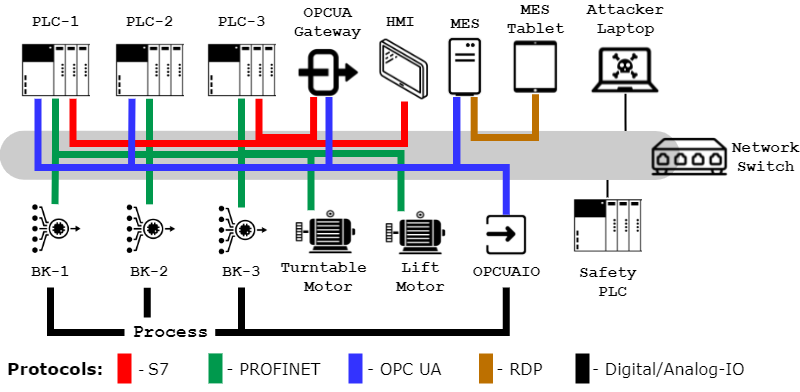}
	\caption{The networking infrastructure of the Festo Demonstrator.}
	\label{fig:festonetworkinfrastructure}
\end{figure}

\noindent \textbf{Network Communication.} The process scenario realized in the \textsl{Festo Demonstrator} is a simplified painting process. It is controlled through network communications between PLCs, I/O devices, actuators and process-associated sensors, and PLCs with Manufacturing Execution Systems (MES) and HMI.
In Figure~\ref{fig:festonetworkinfrastructure}, the network infrastructure of the \textsl{Festo Demonstrator} is shown. 
All the components are connected in STAR topology with the \textsl{Network Switch} at the center. 
The PLCs communicate to bus couplers and motors through \textsl{PROFINET} protocol, whereas PLCs to \textsl{HMI} communication is through \textsl{S7Comm} protocol. 
The process information is relayed to \textsl{MES} through \textsl{OPC UA} protocol from \textsl{OPC UA}-compatible PLCs. 
In case of non \textsl{OPC UA}-compatible \textsl{PLC-3}, an \textsl{OPC UA} gateway collects information from \textsl{PLC-3} through \textsl{S7Comm} and relays it to \textsl{MES}. 
\textsl{RDP} protocol is used to connect a \textsl{Tablet} to \textsl{MES} server to visualize the process execution. 

\subsection{Industrial Network Attack Scenarios}
\label{ssec:networkattackscenarios}

\begin{figure}
	\centering
	\includegraphics[width=0.85\linewidth]{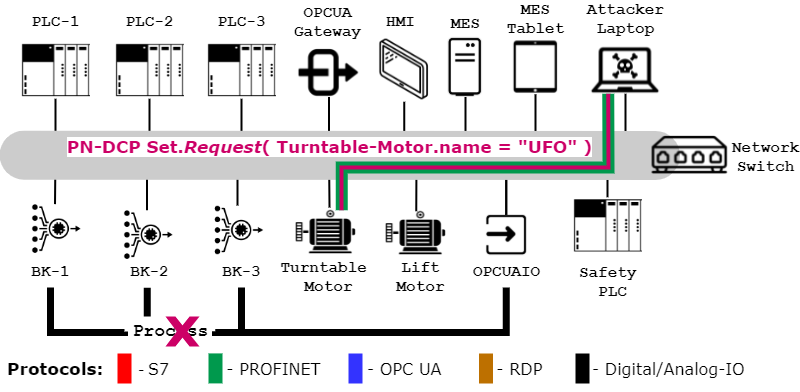}%
	\caption{The Rename Attack on the Festo Demonstrator.}
	\label{fig:festorenameattack}
\end{figure}

An \textit{adversary} is assumed to have gained access to the \textsl{Festo Demonstrator}'s network infrastructure. 
Within \textsl{PROFINET} networks, the components are addressed through their \textit{logical names} for process data exchange via the unencrypted \textsl{PROFINET} protocol.
An \textit{adversary} exploits the \textsl{PROFINET} protocol design flaw and changes the \textit{logical name} of \textsl{Turntable-Motor} to \textit{``ufo''} via the \textsl{PN-DCP} protocol as shown in Figure~\ref{fig:festorenameattack}.
As a result, the other industrial components are not able to identify the component with the name \textit{``Turntable-Motor''} and the process stops. 

%% file: content/03-networkoperations.tex
\section{Network Operation Enumeration for PROFINET}
\label{sec:profinetoperation}

\begin{figure}
	\centering
	\includegraphics[width=0.8\linewidth]{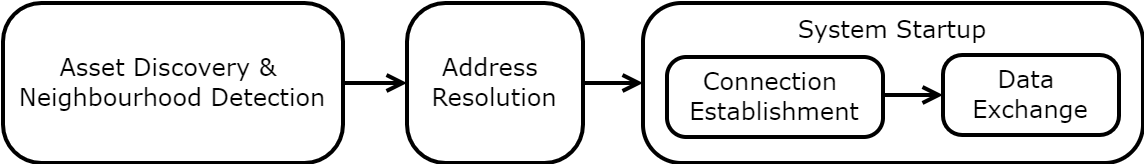}%
	\caption{Networking operations of an industrial system.}
	\label{fig:networkoperations}
\end{figure}

Configuration and commissioning of \textsl{PROFINET}-based automation systems must follow certain mode of operations in a strict order. 
It begins with the System Engineering operation where an automation project is configured in an engineering tool.
General System Description (GSD), an \textsl{XML} file provided by every device manufacturer, contains configuration information for parametrizing the devices for real systems.
In addition, each device is assigned a logical name to address it within the \textsl{PROFINET} communication.
Within the System Engineering mode, an IP address is assigned to each device for communication.
Transmission intervals are defined for cyclic data exchange between controller and devices.
After system engineering is completed, the configuration information is downloaded to the controller.
As soon as the automation system is powered on (or reset), Neighbourhood Detection, Address Resolution and System Startup are the operations followed in the same order, as shown in Figure~\ref{fig:networkoperations}.
With Address Resolution, the controller uses the system configuration information to assign the IP addresses to the devices identified through their pre-assigned logical names.
System Startup operation mode is initiated by the controller to establish connection with devices and configure their I/O parameters.
When the I/O parametrization ends successfully, the controller and devices step into Data Exchange mode to transmit process data, alarms and diagnostic information throughout the network.

Every mode of operation in \textsl{PROFINET}-based automation system, from configuration to commissioning, is accomplished through a complementary network operation involving specific network protocols.
The networking operations and the specific network protocols exchanges driving the operations are summarized. 

\begin{figure}
	\centering
	\includegraphics[width=0.81\linewidth]{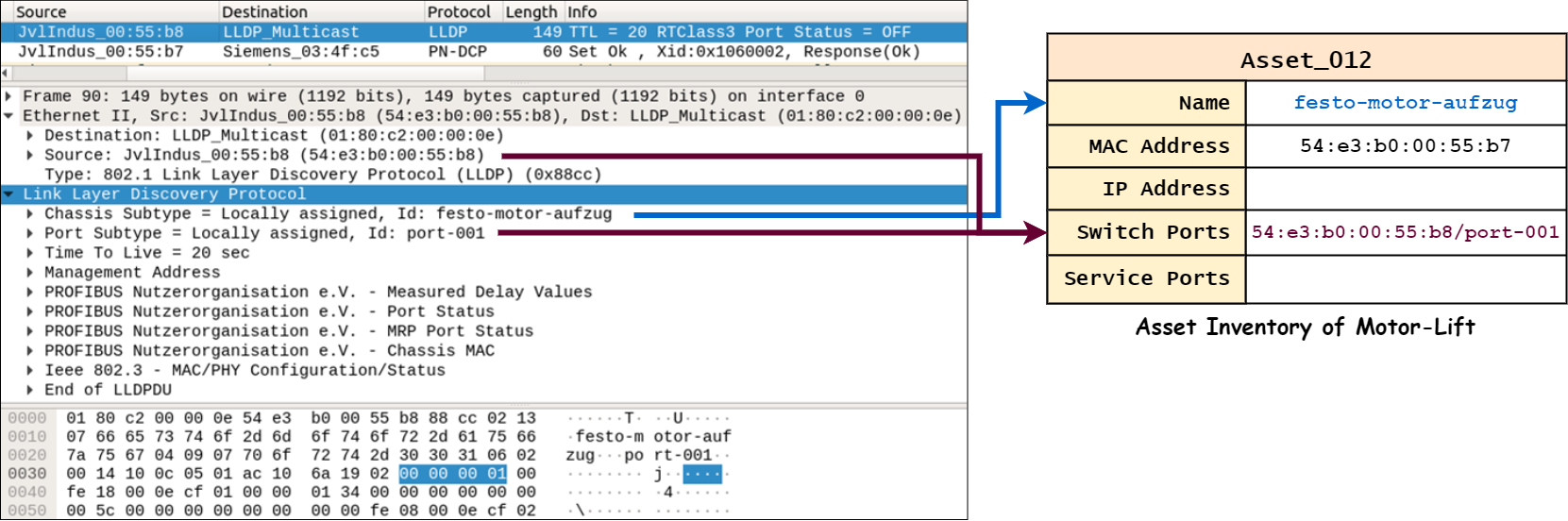}%
	\caption{The demonstration of \textsl{Lift-Motor} device's Asset Inventory filled with information from \textsl{LLDP} frame (\textsl{Wireshark} snippet).}
	\label{fig:lldpinventory}
\end{figure}

\noindent \textbf{Asset Discovery \& Neighbourhood Detection.} 
After the automation system is powered on, the field device's MAC interface and its Physical Device Management (PDev) gets activated to start transmitting the parameters.
PDev contains hardware-level information such as interface name, switch port data, interface and Port MAC addresses and retentively stores IP address and logical name assigned to the device. 
Port information is used by devices to determine their neighbours on port-by-port basis.
Neighbourhood Detection is accomplished through \textsl{Link Layer Discovery Protocol (LLDP)} services. 
\textsl{LLDP}-capable devices communicate with their connected neighbours to cyclically exchange addressing information and consequently determine their physical location. 
\textsl{LLDP} frames are dissected to identify device names, number of switch ports and their MAC addresses for the Automated Asset Inventory, developed for the analysis in the reported work.
Figure~\ref{fig:lldpinventory} shows an example of attributes added for the device \textsl{Lift-Motor} of the \textsl{Festo Demonstrator}. \\

\begin{figure}
	\centering
	\includegraphics[width=0.8\linewidth]{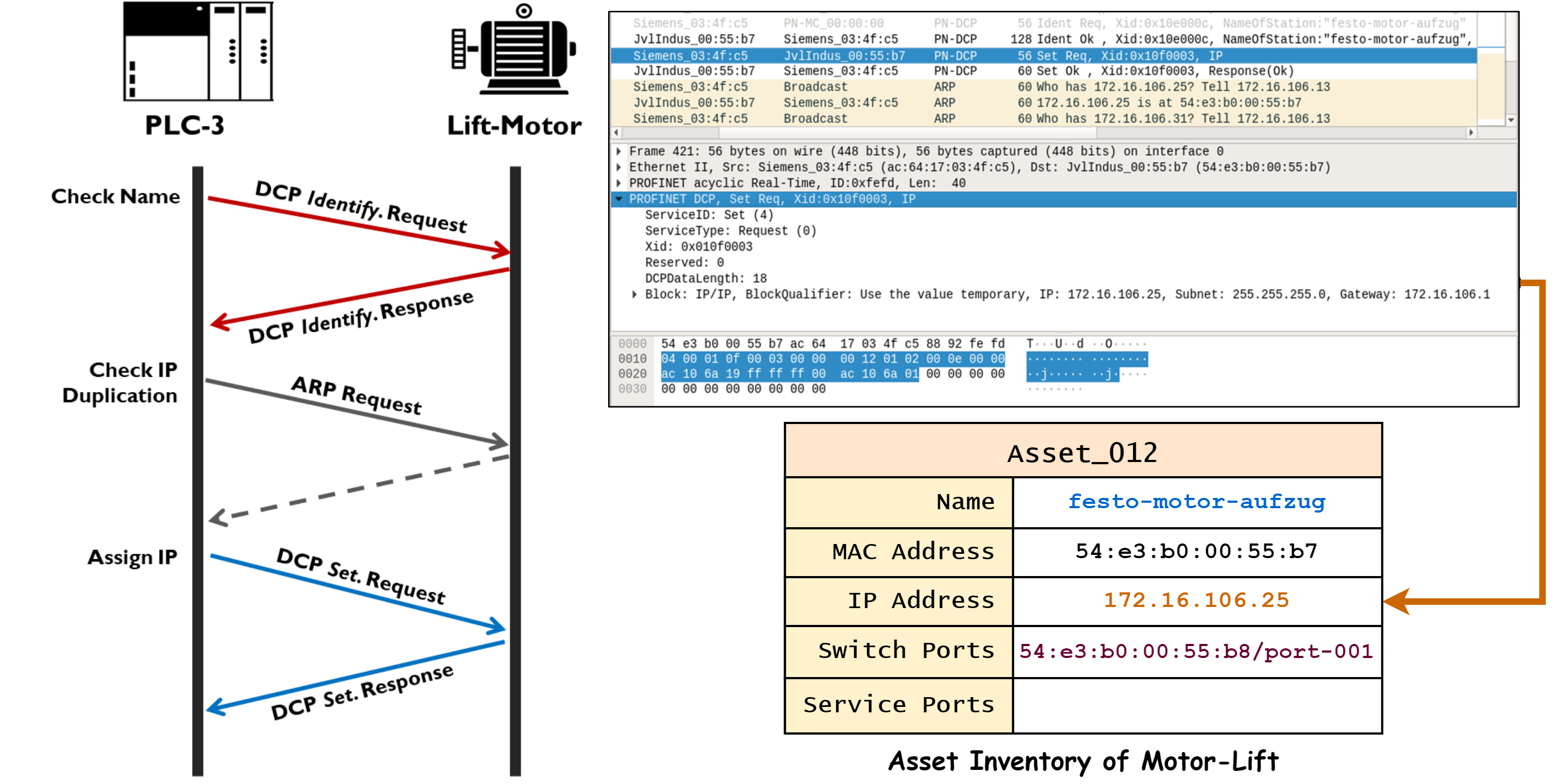}%
	\caption{The demonstration of \textsl{Lift-Motor} device's Asset Inventory from \textsl{PN-DCP} frame.}
	\label{fig:dcpinventory}
\end{figure}

\noindent \textbf{Address Resolution.}
Before the \textsl{PROFINET}-based automation system starts up and field devices start communicating, an IP address needs to be assigned to all devices by the controller.
An IO Device is identified through its `\textit{NameOfStation}' information stored in its PDev and an IP address defined during System Engineering mode is assigned to it. 
Address Resolution networking operation for every device takes place step-by-step as follows: 
\textit{(1)} Controller starts with name resolution and checks for device with configured name through `\textsl{DCP Identify}' service of \textsl{PN-DCP}, 
\textit{(2)} Address Resolution begins with checking if the IP address already exists to avoid assigning same IP address twice through \textsl{ARP}, and 
\textit{(3)} At the end of networking operation, the IP address is assigned to configured device through `\textsl{DCP Set}' service of \textsl{PN-DCP}.
The schematic communication order for Address Resolution between \textsl{PLC-3} and \textsl{Lift-Motor} is shown in Figure~\ref{fig:dcpinventory}.
The information dissected from an \textsl{Address Resolution Protocol (ARP)} \textit{request} packet is used to add the IP address to controller assets.
Information dissected from \textsl{PROFINET Discovery and Configuration Protocol (PN-DCP) Set} is used to add IP address information to configured devices.
The Address Resolution networking operation performed by \textsl{PN-DCP} and \textsl{ARP} execute Address Resolution \textsl{PROFINET} operations. \\

\noindent \textbf{Connection Establishment.}
System Startup operation begins with establishment of communication relationships between the controller and devices via \textsl{PROFINET Context Manager (PN-CM)} protocol communication exchanges.
Through these established communications the controller transmits all the parameters for process data exchange to the internal module of devices.
The process model and associated parameters for devices participating in the process are engineered \& defined during System Engineering operation.

The `\emph{connection}' between an IO Controller and an IO Device is established in an `Application Relationship (AR)' uniquely identified by an \texttt{ARUUID}. 
Within this \textsl{AR}, different `Communication Relationship (CR)' are established for different data exchanges.
An application can access data only through \textsl{CR}s established in an \textsl{AR}.
\textsl{PROFINET} offers \textsl{PN-CM} protocol to handle the Connection Establishment network operation.
The \textsl{PN-CM} network operation uses \textsl{UDP/IP} channel to transmit following frames in the strict order for establishing `\emph{connection}' between controller and device:
\begin{itemize}
	\item \textsl{Connect} frames establish \textsl{AR} and \textsl{CR}s channels.
	\item \textsl{Write} frames parametrize the device \textit{submodule}s.
	\item \textsl{DControl} frames mark the end of parametrization from the controller.
	\item \textsl{CControl} frames mark the validation check of parameters, data structure build up and application readiness from the device.
\end{itemize}

\begin{figure}
	\centering
	\includegraphics[width=0.8\linewidth]{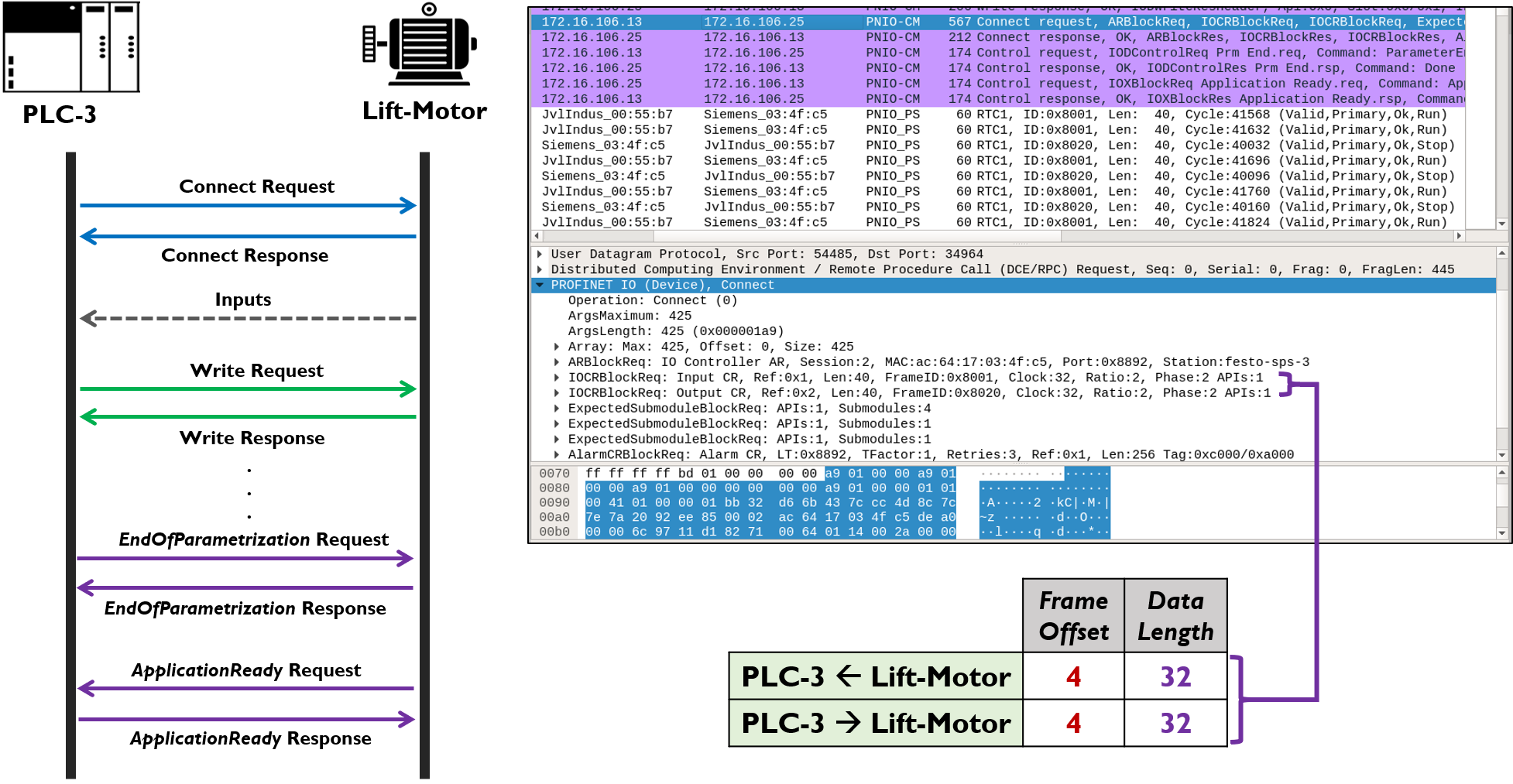}%
	\caption{The Connection Establishment handshake between \textsl{PLC-3} and \textsl{Lift-Motor}.}
	\label{fig:pncm_handshake}
\end{figure}

The first successful exchange of I/O data after \textsl{CControl} frames mark the end of \textsl{PROFINET}'s System Startup operation mode.
The schematic communication order for Communication Establishment between \textsl{PLC-3} and \textsl{Lift-Motor} is shown in Figure~\ref{fig:pncm_handshake}. 
The GSD of a device contains modules and submodules information reflecting the position of process data within the payloads. 
Since in a self-learning analysis from network traffic the GSD isn't accessible, the aforementioned information is extracted through reading and interpreting \textsl{Connect} frames. 
\textsl{PN-CM} frames are dissected to extract the input and output data specifications for device's \textit{submodule}s.
It contains the data format type, order of data (endianness), length of data and the position of data within the payload bytes.

These specifications are used by the Data Exchange network operation's \textsl{PROFINET Input/Output (PNIO)} frames to extract process bytes.
The `\emph{connection}' between controller and device following a Connection Establishment network operation guides building the logical network topology of the system.
These separated ISO layer connection state information between network assets is maintained throughout the automation system's runtime and deviations are reported. \\

\begin{figure}
	\centering
	\includegraphics[width=0.85\linewidth]{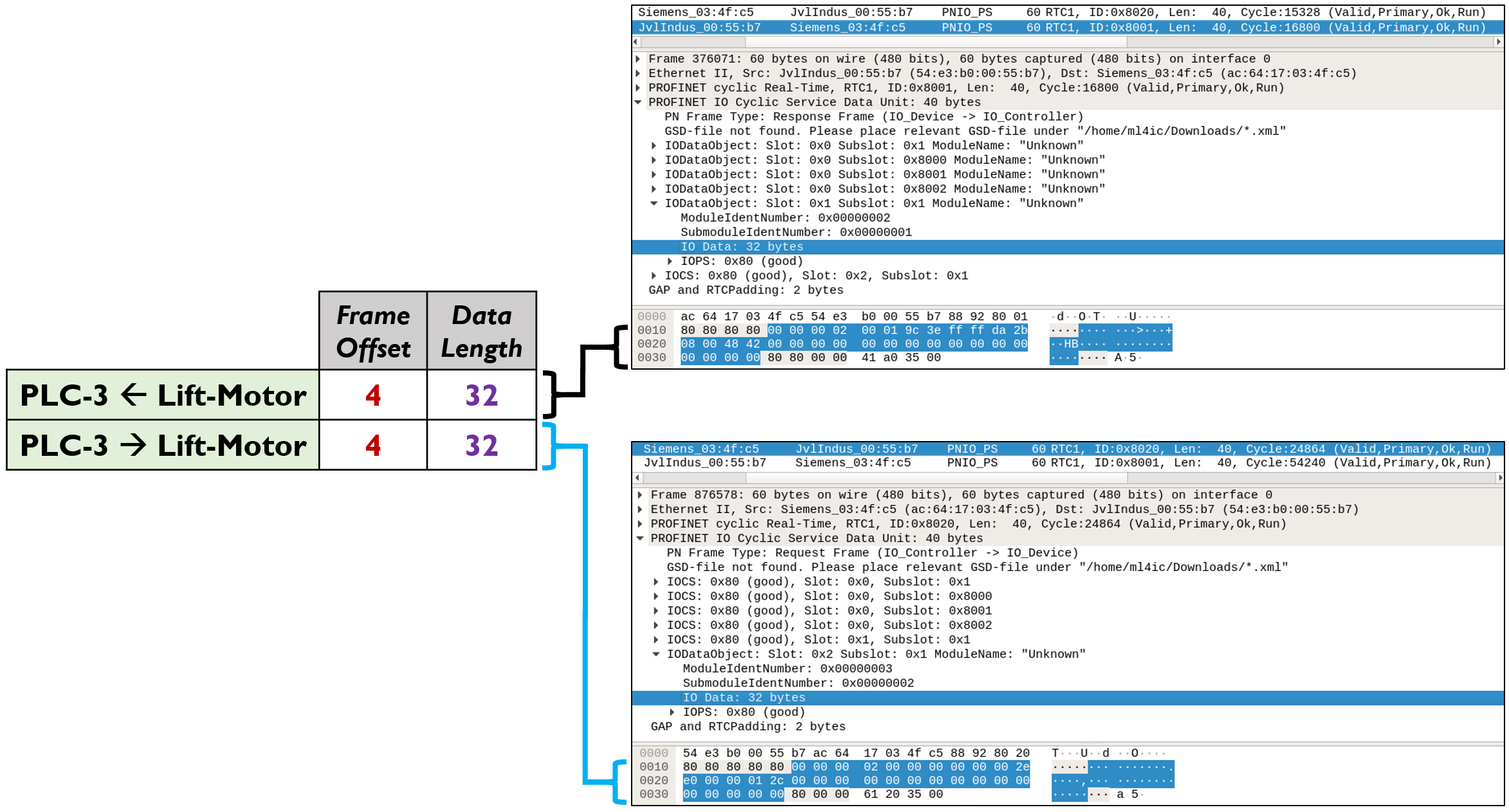}%
	\caption{The demonstration of extracting process data between \textsl{PLC-3} and \textsl{Lift-Motor} from \textsl{PNIO} frames  (\textsl{Wireshark} snippet), using parameters extracted from \textsl{Connect} frame.}
	\label{fig:pnioextraction}
\end{figure}

\noindent \textbf{Data Exchange.}
Once the System Startup operation establishes \textsl{AR} and data specific \textsl{CR}s between devices and controller, the connection-oriented communication channel is set for exchange of cyclic process data, acyclic diagnostic data and alarms.
\textsl{PNIO} protocol defines the format and context for data exchange.
Cyclic \textsl{PNIO} frames are sent unacknowledged between controller and devices.
After \textsl{CControl} frames are acknowledged by the controller, the first valid exchange of I/O data with \texttt{IOPS}=\texttt{GOOD} ends Connection Establishment operation.
Data Exchange operation begins with cyclical exchange of process data at configured/parametrized fixed intervals.
\textsl{PNIO} cyclic data frame is transmitted in real-time with \texttt{Ethertype}=\texttt{0x8892}
The process data bytes from \texttt{Data} field are extracted using information from \textsl{Connect} frame dissection, as shown in Figure~\ref{fig:pnioextraction}. \\

Enumeration of \textsl{PROFINET} operation modes is performed by passively monitoring/analysing the network traffic and identifying the associated network operation stage-by-stage. 
System Engineering mode is performed offline, hence, it can't be enumerated through analysing the network traffic. 

%% file: content/04-poet.tex
\section{PROFINET Operations Enumeration and Tracking}
\label{sec:enumerateandtrack}

Through monitoring an industrial system's communication network, its characteristics are observed to build normal operations behaviour baseline.
Deviations from the baseline behaviour could be triggered by \textit{physical} or \textit{cyber} threats. 
A systematic framework is needed to enumerate operations extracted from network traffic and track them to report deviations.

\textsl{PROFINET}-based automation systems follow strict order of operations.
The operation mode and corresponding network operations with associated network protocols have been outlined in Section~\ref{sec:profinetoperation}.
All of those network operation's dissected information are combined to systematically iterate over the \textsl{PROFINET} operation modes as and when there occurrences are observed through network analysis.	
\textsl{PROFINET} devices transit through \textsl{PROFINET} operations to establish \emph{connection}s between them for cyclic and acyclic data exchange. 
These transitions also govern transitions in \textsl{PROFINET} connections, which constitutes the logical topology and industrial process behaviour of the \textsl{PROFINET} system. 

Finite State Machines(FSM) \cite{kleene1956representation} are widely used for protocol specification (e.g. \textsl{TCP/IP} \cite{stevens1993tcp}) \cite{holzmann1993design}, where the valid \textit{transitions} and \textit{states} of message exchanges are defined. 
For the requirements of capturing the transitions between industrial operations triggered by the communication events, the FSMs are modelled to enumerate the \textsl{PROFINET} operations of the device, connection and system. 
\textsl{PROFINET} standard, the informative handbook on \textsl{PROFINET} \cite{popp} and empirical information collected from analysing real-world \textsl{PROFINET}-based system communications are interpreted to model the operations in FSMs. 

In the next subsections, each FSM is described with the overview of states and events triggering the transitions outlined in its state diagram. 
The transitions which are modelled based on empirical information are distinguished by dashed edges and details are presented. 

\subsection{FSM \textsl{PROFINET} Device}

\begin{figure}
	\centering
	\includegraphics[width=\textwidth]{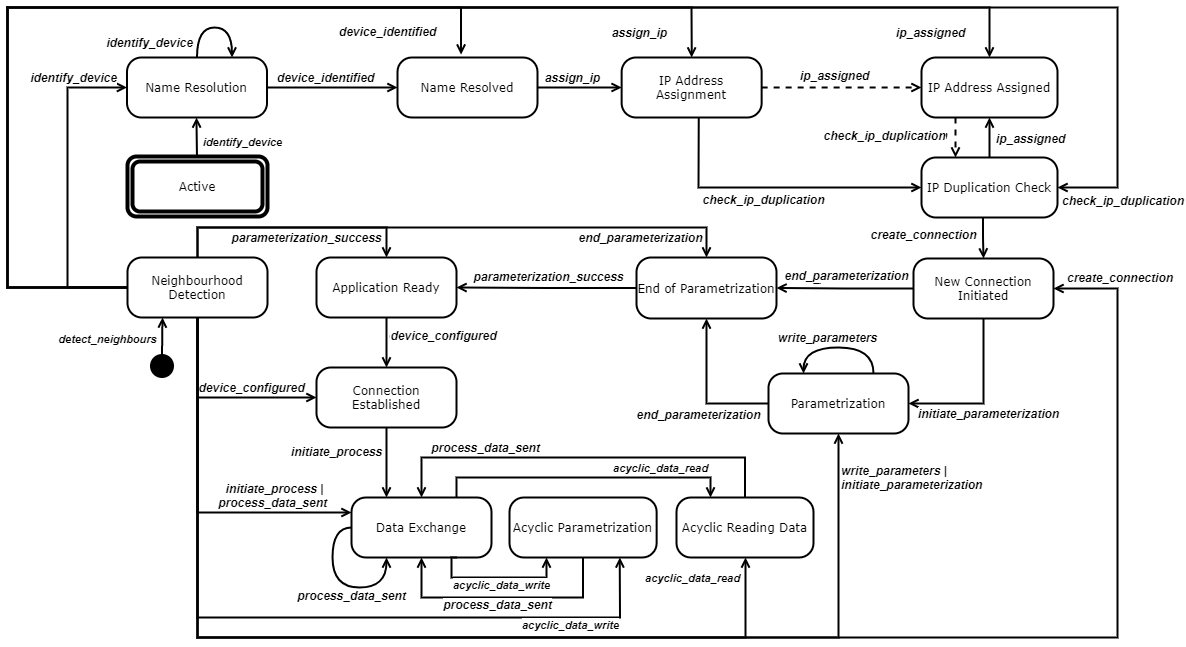}%
	\caption{\textsl{PROFINET} Device State Machine.}
	\label{fig:stmdevice}
\end{figure}

\noindent  \textbf{States and transitions.}
FSM Device enters with \emph{Active} state as soon as the system is powered on, shown in Figure~\ref{fig:stmdevice}. 
It transits either to \emph{Neighbourhood Detection} or \emph{Name Resolution} state depending on the event triggered.
FSM follows through the transitions as and when the triggering event is detected in network traffic.

\textsl{LLDP} frames are periodically transmitted by \textsl{PROFINET} device as per their \texttt{Time-To-Live} value for consistent \textsl{LLDP} information validation.
Hence, \emph{Neighbourhood Detection} state can be arrived from any other state whenever \textit{detect\_neighbours} event is triggered by \textsl{LLDP} frame.
Consequently, all the states are reachable with corresponding triggering events from \emph{Neighbourhood Detection} state.

Through network traffic analysis of \textsl{PROFINET}-based systems with Siemens PLC, transitions - \emph{IP Address Assignment} to \emph{IP Address Assigned} and \emph{IP Address Assigned} to \emph{IP Duplication Check} - have been modelled. 
Deviating from transitions mentioned in the literature, the \textsl{PROFINET} devices checked for IP duplication with \textsl{Gratuitous ARP} after the IP address has been assigned to them. 
These transitions are also verified on different \textsl{PROFINET}-based systems. \\

\noindent \textbf{Relationship between states and \textsl{PROFINET} operations.}
\emph{Neighbourhood Detection} state constitutes Asset Discovery \& Neighbourhood Detection \textsl{PROFINET} operation. 
States \emph{Name Resolution}, \emph{Name Resolved}, \emph{IP Address Assignment}, \emph{IP Address Assigned} and \emph{IP Duplication Check} constitute Address Resolution \textsl{PROFINET} operation.
\textsl{PROFINET}'s Connection Establishment operation consists of states \emph{New Connection Initiated}, \emph{Parametrization}, \emph{End Of Parametrization}, \emph{Application Ready} and \emph{Connection Established}.
States \emph{Data Exchange}, \emph{Acyclic Parametrization} and \emph{Acyclic Reading Data} reflect Data Exchange \textsl{PROFINET} operation.

\subsection{FSM \textsl{PROFINET} Connection}

\begin{figure}
	\centering
	\includegraphics[width=0.85\linewidth]{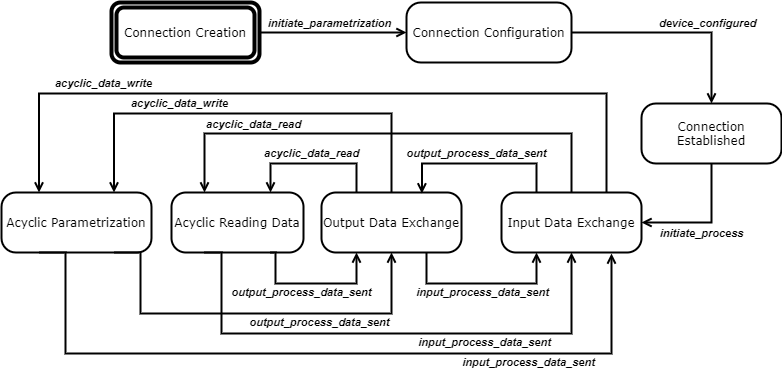}%
	\caption{\textsl{PROFINET} Connection State Machine.}
	\label{fig:stmconnection}
\end{figure}

\textbf{States and transitions.}
A \textsl{PROFINET} connection is established between \textsl{PROFINET} devices through \textsl{PROFINET}'s \textsl{PN-CM} protocol handshake.
The cyclic and acyclic data exchange takes place through this connection. 
Hence, each connection is identified through MAC addresses of participating \textsl{PROFINET} devices.
FSM Connection enters with \emph{Connection Creation} state as soon as \textsl{Connect} \textit{request} frame is sent by the controller, shown in Figure~\ref{fig:stmconnection}. 
FSM follows through the transitions as and when the triggering event is detected in network traffic.
In particular, events \textit{output\_process\_data\_sent} and \textit{input\_process\_data\_sent} are triggered by transmission of \textsl{PNIO} frames from controller to device and vice versa, respectively. \\

\noindent \textbf{Relationship between states and \textsl{PROFINET} operations.}
States \emph{Connection Creation}, \emph{Connection Configuration} and \emph{Connection Established} constitute Connection Establishment \textsl{PROFINET} operation.
\textsl{PROFINET}'s Data Exchange operation are reflected in states \emph{Input Data Exchange}, \emph{Output Data Exchange}, \emph{Acyclic Parametrization} and \emph{Acyclic Reading Data}.

\subsection{FSM \textsl{PROFINET} System}

\begin{figure}
	\centering
	\includegraphics[width=0.8\linewidth]{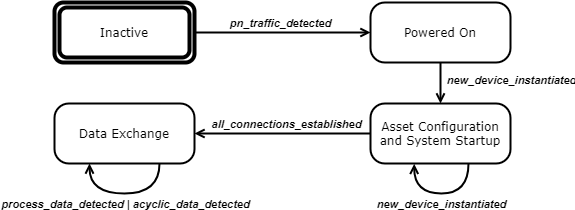}%
	\caption{\textsl{PROFINET} System State Machine.}
	\label{fig:stmsystem}
\end{figure}

\textbf{States and transitions.}
FSM System initializes with \emph{Inactive} state and transits into \emph{Powered On} as soon as \textsl{PROFINET} traffic triggers event \textit{pn\_traffic\_detected}, show in Figure~\ref{fig:stmsystem}.
FSM follows through the transitions as and when the triggering event is detected in network traffic.
Event \textit{all\_connections\_established} is triggered when all the FSM \textsl{PROFINET} Connection instances have arrived in state \emph{Connection Established}. \\

\noindent \textbf{Relationship between states and \textsl{PROFINET} operations.}
State \emph{Powered On} reflects either \textsl{PROFINET} operation Asset Discovery \& Neighbourhood Detection or Address Resolution if the event \textit{pn\_traffic\_detected} is triggered by \textsl{LLDP} or \textsl{DCP Identify} \textit{request} frames, respectively.
State \emph{Asset Configuration \& System Startup} reflects Connection Establishment \textsl{PROFINET} operation.
\textsl{PROFINET}'s Data Exchange operation of cyclic and acyclic data transmission is reflected in state \emph{Data Exchange}.

\subsection{Framework}

\textsl{PROFINET} system, connection and device FSMs are realized in a \textit{python}-based framework, termed as PROFINET Operations Enumeration and Tracking (POET), for enumeration and tracking \textsl{PROFINET}-based industrial network communication.
It is implemented with \textit{pytransitions} \cite{pytransitions} and logs transitions in FSM instances of \textsl{PROFINET} System, Connection and Device continuously.
Each FSM \textsl{PROFINET} System instance is identified by a name given while initialization, whereas the device name extracted from network traffic (\textsl{DCP Identify} \textit{request}/\textsl{LLDP} frame) is used for identifying FSM \textsl{PROFINET} Device instance.
FSM \textsl{PROFINET} Connection instance is identified by the connection identifier created from concatenating MAC addresses of devices.

POET clearly satisfies the BSI's general requirements category for an anomaly detection system to identify communicating devices, protocols and communication links (modeled as \textsl{PROFINET} Connection) in the industrial network.

%% file: content/05-anomalydetection.tex
\section{Anomaly Detection with POET}
\label{sec:poetusage}

Industrial networks are vulnerable to different threat behaviours, each utilizing different techniques to exploit industrial network characteristics. 
\emph{MITRE ATT\&CK for ICS} \cite{mitre} is a knowledge base of such industrial system targeted threat behaviours, collected through cyber threat intelligence reports of known cyber incidents.
Some threat behaviours (such as \emph{Modify Parameter}, \emph{Denial of Service}) are targeted at the industrial network operation to disrupt the underlying industrial process.

\begin{figure}
	\centering
	\includegraphics[width=\linewidth]{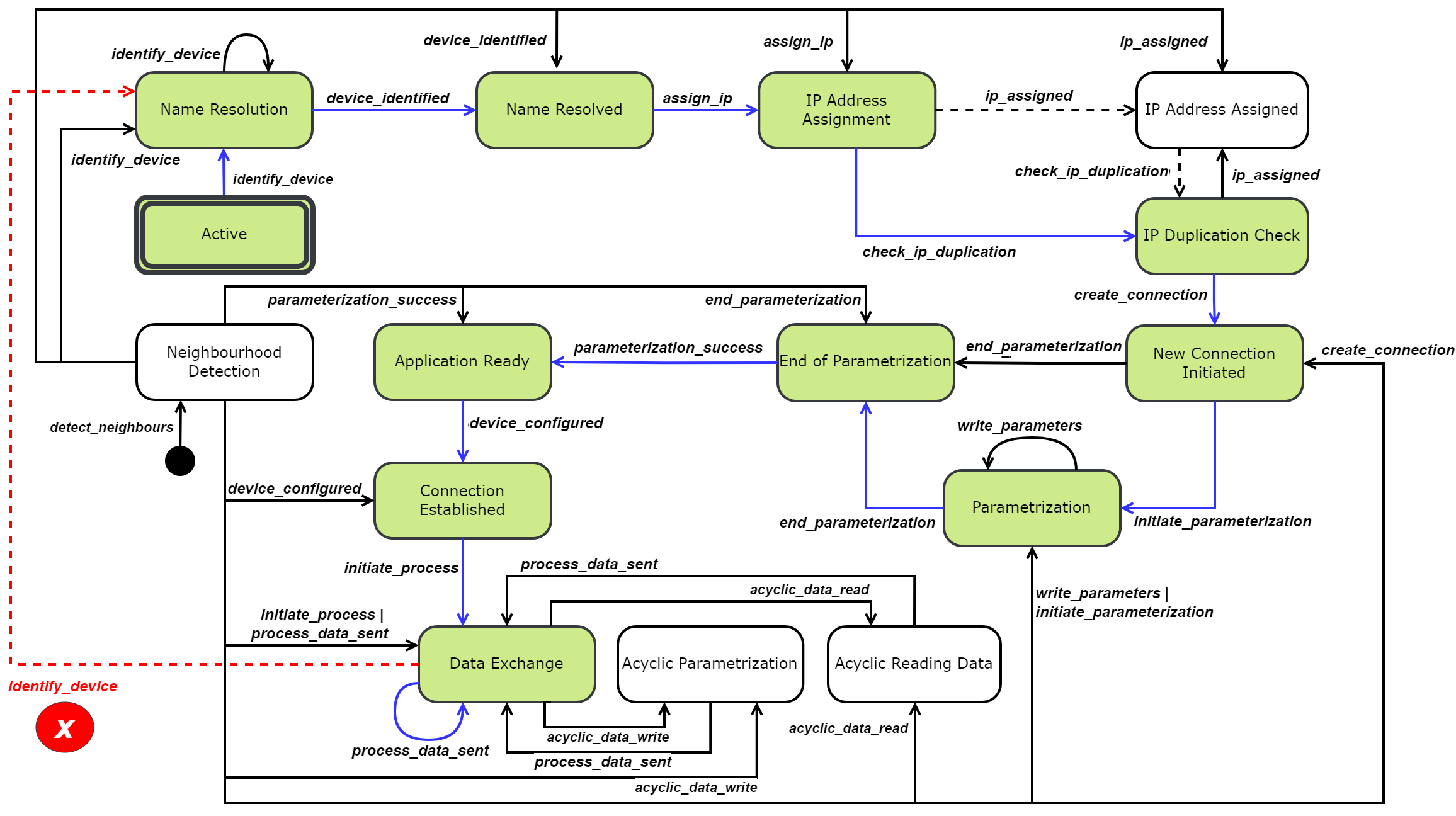}%
	\caption{The detection of Rename Attack on the \textsl{Festo Demonstrator} with the POET.}
	\label{fig:poetlogger}
\end{figure}

Pfrang et. al. \cite{pfrang} outlined threat scenarios targeted at real-world \textsl{PROFINET}-based systems with two different techniques to take over control of a \textsl{PROFINET} device.
Within \textsl{PROFINET} networks, the devices are identified through the assigned logical name for process data exchange. 
The first attack of \cite{pfrang} demonstrated how an attacker changes the name of a device utilizing \textsl{PN-DCP} protocol and disconnecting it with other devices.
A similar attack, the `\textit{Rename Attack}' was performed on the \textsl{Festo Demonstrator} as outlined in Section~\ref{ssec:networkattackscenarios}  and POET was employed to monitor network traffic.
The attack triggered events which aren't allowed for FSM Device instance of `festo-motor-scheibe' and were reported in POET's logger, as shown in Figure~\ref{fig:poetlogger}.
The blue edges are the \textit{valid} \textsl{PROFINET} operation transitions triggered by the appropriate protocol events. 
As per the \textsl{PN-DCP} protocol specification, the packets (\textit{x}) are \textit{valid}, however, they violate \textsl{PROFINET} device's \textit{valid} operation transition represented as red edges, and thus detected.

The second attack of Pfrang et. al. \cite{pfrang} is essentially disrupting \textsl{PROFINET} network operations by establishing new connection with device. 
This action initiates Connection Establishment \textsl{PROFINET} operation which isn't valid transition state for FSM Device instance.
Employing POET in such scenarios would also detect this attack and enhance visibility to unwarranted events in \textsl{PROFINET} networks with explanation.

POET's capability to detect the two attacks outlined in \cite{pfrang} satisfies the BSI's category of requirements to detect  unusual or exceptional activities in an ICS network.
Any other attacks that would violate the validity of an industrial operation through protocols other than \textsl{PN-DCP} and \textsl{PN-CM} would be detected by POET. 

\subsection{Discussion}

Protocol-analysis based IDS have been proposed for industrial protocols such as \textsl{DNP3}, \textsl{Modbus/TCP}, \textsl{GOOSE}, etc.~\cite{icsids}, where protocol specifications are utilized to build system profile and the deviations are reported. 
The proposed FSM-based framework, POET, can be categorized along with them as Protocol-analysis based IDS for \textsl{PROFINET}. 

POET is the first-of-its-kind Protocol-analysis based IDS for \textsl{PROFINET} to incorporate empirical behaviour of \textsl{PROFINET} system collected from real-world systems.
The current version of POET uses empirical information collected from \textsl{PROFINET} systems incorporating Siemens PLC, it could vary with other PLC environments (e.g. CODESYS \cite{codesys}), devices and can be adapted. 
The extraction of network events triggering the transitions of industrial operations would be adapted to the new protocol communication stack. 

POET offers insights into the operations of \textsl{PROFINET}-based systems at ascending granularity of system, connection between devices and device. 
This granularity helps to specify events proficiently to be used in downstream analysis for anomaly detection. 
For example, Pfrang et. al. \cite{pfrang} outlined enhanced Snort for \textsl{PROFINET} which was used to detect the two attacks mentioned in their work.
It can be integrated with POET to trigger alarms when unwarranted transitions occur.

At the end, we demonstrated a successful workflow to interpret an industrial protocol specification and the empirical information collected from its real-world industrial system usage to design a Protocol-analysis based IDS. 
A similar workflow could be utilized for another industrial protocol such as \textsl{EtherNet/IP} utilizing the outlined FSM models at different granularities of system, connection and device. 
The FSM models would have to be adapted for states and triggering events.  

%% file: content/06-conclusion.tex
\section{Conclusion}
\label{sec:conclusion}

The \textsl{PROFINET} network traffic is mapped to different \textsl{PROFINET} operations for interpreting the underlying status of industrial communication.
In Section~\ref{sec:profinetoperation} the solution to \textbf{Problem 1} is outlined, where different protocols associated with \textsl{PROFINET} operations are mapped to \textsl{PROFINET} network operations.
For every network operation, the role it plays within \textsl{PROFINET}-based automation system communication and the network protocol utilized to achieve the goal has been presented. 
Thus, satisfying the BSI's general requirements for network transparency.

In addition, the protocol associated communication behaviour and their detection through protocol frame analysis has been outlined. 
As the solution to \textbf{Problem 2}, Section~\ref{sec:enumerateandtrack} modelled operations of \textsl{PROFINET} system, connection and device as  Finite State Machines (FSM) to systematically enumerate and track \textsl{PROFINET} operations.
\textsl{PROFINET} Device, Connection and System FSMs are realized in a \textit{python}-based framework named PROFINET Operations Enumeration and Tracking (POET).    
Its successful usage as Protocol-based IDS in detecting cyber attack on real-world \textsl{PROFINET} demonstrator has been presented in Section~\ref{sec:poetusage}. 
This demonstrates POET as an anomaly detection solution that satisfies the BSI's requirement to identify unusual or exceptional activities in an ICS network.
In conclusion, the challenge of self-learning \textsl{PROFINET}-based industrial communication networks is solved through interpretation of network traffic to \textsl{PROFINET} operations.
The workflow developed to interpret an industrial protocol's specification with the empirical information from its real-world usage to develop an anomaly detection system can be replicated further to other industrial networking technology. 